\documentstyle[twocolumn,aps,prl,graphics,epsfig]{revtex}
\begin{document}

\title{Ultra-bright source of polarization-entangled photons}
\author{Paul G. Kwiat$^1$, Edo Waks$^1$\footnotemark, 
Andrew G. White$^1$, Ian Appelbaum$^1$\footnotemark,
and Philippe H. Eberhard$^2$}
\address{$^1$Physics Division, P-23, Los Alamos National Laboratory,
Los Alamos, New Mexico 87545}
\address{$^2$Lawrence Berkeley Laboratory, University of California,
Berkeley, California  94720}

\date{Resubmitted to PRA Apr. 19, 1999, shortened May 17, 1999}
\draft
\maketitle

\begin{abstract}
Using the process of spontaneous parametric down-conversion in a 
novel two-crystal geometry, we have generated a source of
polarization-entangled 
photon pairs which is more than ten times brighter, 
per unit of pump power, than
previous sources, with another factor of 30 to 75 expected to be readily
achievable.  We have measured a high level of entanglement 
between photons emitted over a relatively large collection angle, 
and over a 10-nm bandwidth.  As a demonstration of the 
source capabilities, we obtained a 242-$\sigma$ violation of Bell's 
inequalities in less than three minutes, and 
observed near-perfect photon correlations
when the collection efficiency was reduced.  In addition, both the
degree of entanglement and the state purity should be readily
tunable.
\end{abstract}

\pacs{PACS numbers: 03.65.Bz,42.50.Dv, 03.67.-a}

Entangled states of multiparticle systems are arguably the 
quintessential feature of quantum mechanics \cite{Schroedinger-quote}.  
In addition to their central role in discussions of 
nonlocal quantum correlations 
\cite{EPR}, they form the basis of 
quantum information, and enable such phenomena as quantum 
cryptography \cite{crypt}, dense coding \cite{dense}, 
teleportation \cite{teleportation}, and quantum 
computation \cite{qcomputing}. At present, the most accessible 
and controllable source of entanglement arises from 
the process of spontaneous parametric 
down-conversion in a nonlinear optical crystal.  Here we describe a 
proposal for, and experimental realization of, an ultrabright source 
of polarization-entangled photon pairs, using {\it two} 
such nonlinear crystals. Because 
nearly every pair of photons produced is polarization-entangled, 
the total flux of emitted polarization-entangled pairs should be
hundreds of times greater than is achievable with the best 
previous source, for comparable pump powers.  The new technique
has the added advantage that the degree 
of entanglement and the purity of the state may be readily
tunable, heretofore impossible. 

It is now well known that the photons produced via the 
down-conversion process
share nonclassical correlations \cite{review}.  In particular, when a 
pump photon splits into two daughter photons, conservation of energy 
and momentum lead to entanglements in these two continuous degrees of 
freedom \cite{Franson}.  
Yet conceptually, the simplest examples of entangled 
states of two photons are the {\it polarization}-entangled ``Bell 
states":
\begin{equation}
|H_1,V_2\rangle\pm |V_1,H_2\rangle\;\; ;
|H_1,H_2\rangle\pm |V_1,V_2\rangle\;\, ,
\end{equation}
where $H$ and $V$ denote horizontal and vertical polarization, 
respectively, and for convenience we omit the normalization factor
($1/\sqrt{2}$).  
For instance, $HV-VH$ is the direct analog of the 
spin-singlet considered by Bell \cite{EPR}.  To date there have been 
only two methods for producing such polarization-entangled 
photon pairs, and each has fairly substantial limitations. The first 
was an atomic cascade -- a two-photon decay process 
from one state of zero angular momentum to another. The resulting 
photons {\it do} display nonclassical correlations (they were used 
in the first tests of Bell's inequalities
\cite{Clauser,Aspect}), but the correlations decrease if the photons 
are not emitted back-to-back, as is allowed by recoil of the 
parent atom.  

This problem was circumvented with parametric down-conversion, 
since the emission directions of the photons are well-correlated. 
In several earlier experiments down-conversion photon pairs of 
{\it definite} polarization were incident on a beamsplitter, and nonclassical
correlations observed for those {\it post-selected} events in 
which photons traveled to different output ports \cite{postselection}.  
However, the photons were actually created in polarization 
{\it product}-states.

A source of truly polarization-entangled photons was realized using 
down-conversion with type-II phase-matching, in which the 
photons are produced with (definite) orthogonal 
polarizations  \cite{BBOtype2}.  
For two particular emission directions, however, the correlated photons 
are produced in the state $HV+VH$; additional birefringent elements in 
one or both beams allow the formation of all four  
Bell states.  This source has been
employed to demonstrate quantum dense coding \cite{dense-expt}, 
teleportation \cite{tele-expts}, a post-selection-free test 
of Bell's inequality for energy and time variables \cite{Strekalov}, a 
test of Bell's inequality (for polarization variables) free of the usual 
rapid-switching loophole \cite{Belltest}, and most recently, the 
generation of GHZ states of {\it three} photons \cite{GHZ}. 
Coincidence count rates of up to $\sim 2000$s$^{-1}$ (for a 3-mm
thick BBO crystal and a 150mW pump)  have been 
observed with this source, while maintaining an acceptable 
level of entanglement.

Nevertheless, the source brightness is still very limited because 
the photons are polarization-entangled only along two special 
directions.  Using a {\it two}-crystal geometry, 
we have constructed a source in which {\it all} pairs of a given color 
are entangled, and we expect that this should extend to most, if not 
all, of the spectral down-conversion output, i.e., to cones 
corresponding to different colors \cite{twocrystal}. 
Consider two adjacent, relatively thin, nonlinear crystals, 
operated with type-I phase-matching (Fig. 1a).  
The identically-cut crystals are oriented with their optic axes aligned 
in perpendicular planes, i.e., the first (second) crystal's optic axis 
and the pump beam define the
\begin{figure}
\begin{center}
\epsfxsize=\columnwidth
\epsfbox{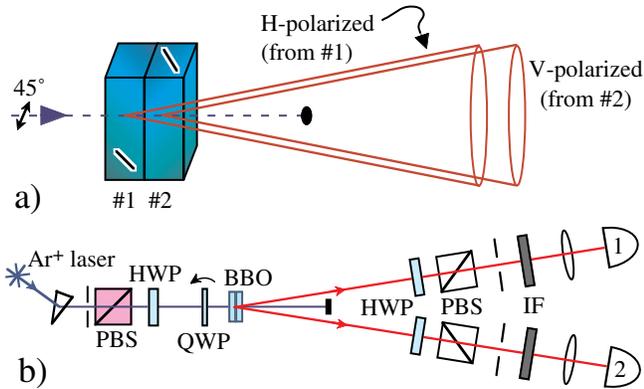}
\end{center}
\caption{a) Method to produce polarization-entangled photons from 
two identical down-conversion crystals, oriented at $90^{\circ}$ 
with respect to each other, 
i.e., the optic axis of the first (second) lies in the 
vertical (horizontal) plane. 
b) Experimental setup 
to pump and characterize the source.}
\label{fig_1}
\end{figure}
\noindent vertical (horizontal) plane.  With a vertically polarized pump beam, 
due to the type-I coupling, down-conversion will only occur in crystal 
1, where the pump is extraordinary polarized -- the resulting 
down-conversion light cones will be horizontally polarized.  
Similarly, with a horizontally-polarized pump, down-conversion will 
only occur in the second crystal, producing otherwise identical cones 
of vertically-polarized photon pairs.  A $45^{\circ}$-polarized pump 
photon will be equally likely to down-convert in either crystal 
(neglecting losses from passing through the first), and these two 
possible down-conversion processes are {\it coherent} with one 
another, as long as the emitted spatial modes for a given pair of 
photons are indistinguishable for the two crystals \cite{overlap}.  
Consequently, the photons will automatically be created in the state 
$HH + e^{i \phi} VV$.  $\phi$ is determined by the details of the 
phase-matching and the crystal thickness, but can be adjusted by 
tilting the BBO crystals themselves (but this changes the cones' 
opening angles), by imposing a birefringent phase shift on one of the 
output beams, or by controlling the relative phase between the 
horizontal and vertical components of the {\it pump} light.

Figure 1b shows the experimental setup used to produce and 
characterize the correlated photons.  The \mbox{$\sim 2$mm}-diameter 
pump beam at 351.1nm was produced by an Ar$^{+}$ laser, and directed 
to the two crystals after passing through: a dispersion prism to 
remove unwanted background laser fluorescence; a polarizing 
beamsplitter (PBS) to give a pure polarization state; a rotatable half 
waveplate (HWP) to adjust the angle of the linear polarization; and a 
second, tiltable waveplate for adjusting $\phi$.  The nonlinear 
crystals themselves were BBO (8.0 x 8.0 x 0.59 mm), optic axis cut at 
$\theta_{pm} = 33.9^{\circ}$.  For this cut the degenerate-frequency 
photons at 702nm are emitted into a cone of half-opening angle 
$3.0^{\circ}$.  For most of the data presented here, interference 
filters (IF) centered at
\begin{figure}
\begin{center}
\epsfxsize=\columnwidth
\epsfbox{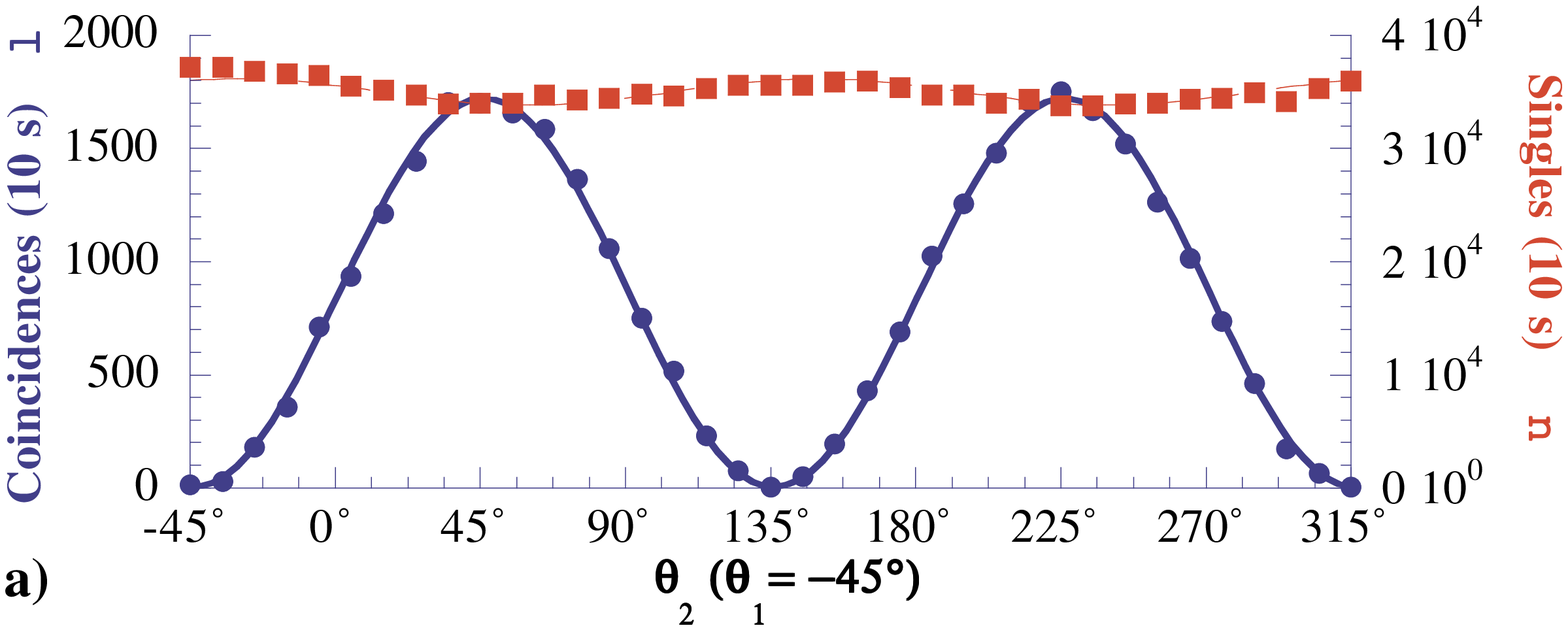}
\epsfxsize=\columnwidth
\epsfbox{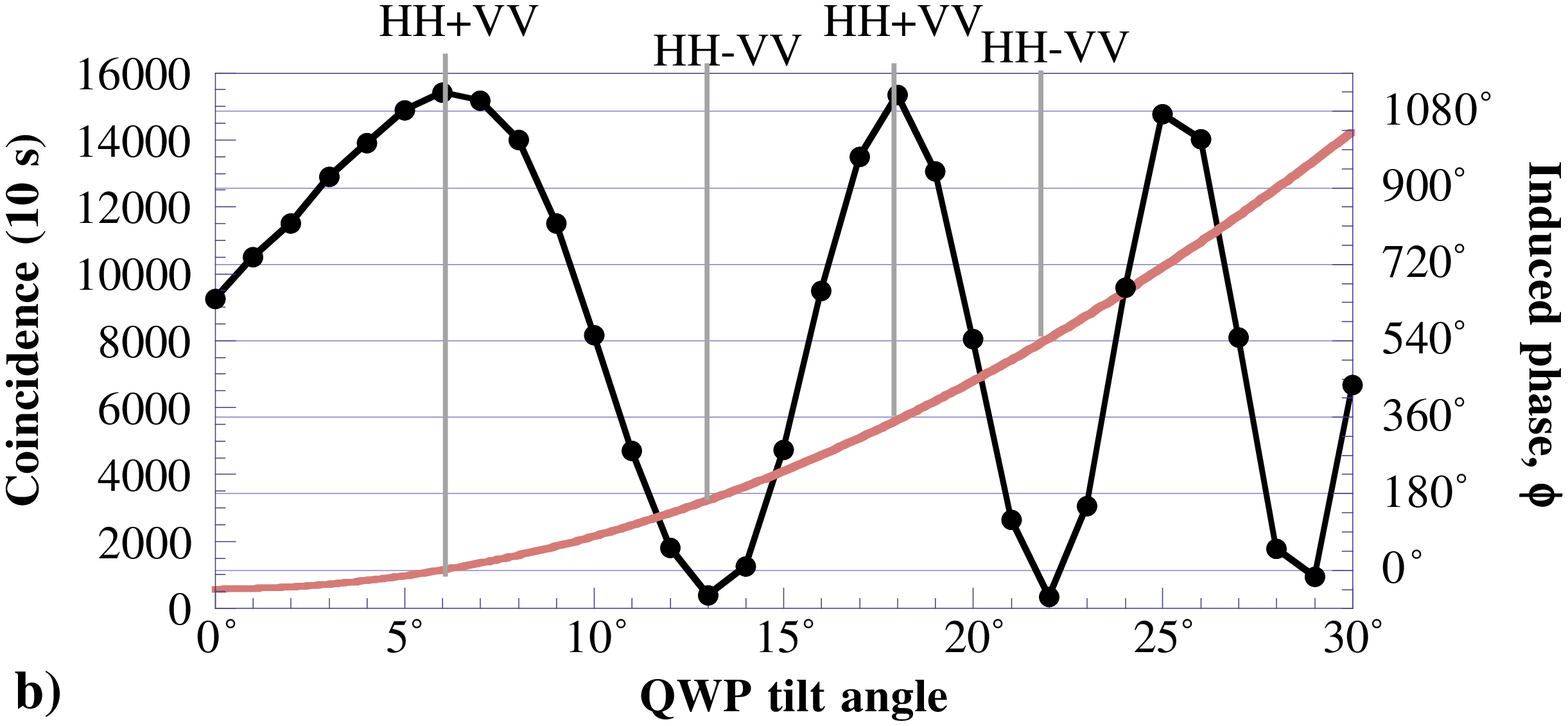}
\end{center}
\caption{a) Measurements of the polarization 
entanglement. The polarization analysis of photon 2 
was varied, while that of photon 1 was at $-45^{\circ}$.  The rate at
detector 2 (squares, right axis) is essentially 
constant, i.e., the photons are individually nearly unpolarized, while the 
coincidence rate (circles, left axis) displays the expected quantum
mechanical correlations. The solid curve is a best fit, with
visibility $V = 99.6 \pm 0.3\%$. 
b)~Coincidences as the relative phase $\phi$ was varied by tilting the 
waveplate just before the crystal; both photons were analyzed at 
$45^{\circ}$.  The solid curve is the calculated phase shift for our 2-mm 
thick zero-order quartz quarter waveplate, adjusted for the residual phase
shift from the BBO crystals themselves.}
\label{fig_2}
\end{figure}
\noindent 702nm (FWHM $\approx 5$nm) were used to reduce background and select 
only these (nearly-)degenerate photons; the maximum transmission of 
these filters was $\sim~65\%$.

The polarization correlations were measured using adjustable 
polarization analyzers, each consisting of a PBS preceded by 
an adjustable HWP (for 702nm). After passing through adjustable 
irises, the light was collected using 
35mm-focal length doublet lenses, and 
directed onto single-photon detectors ---
silicon avalanche photodiodes (EG\&G \#SPCM's), 
with efficiencies of \mbox{$\sim 65\%$} and dark count rates of order
100s$^{-1}$. The outputs of the detectors were recorded 
directly (``singles'') and in coincidence, using a time to amplitude 
convertor and single-channel analyzer.  A time window of 7 ns 
was found sufficient to capture the true coincidences.
Typical ``accidental'' coincidence rates 
were negligible ($<1$s$^{-1}$).

Figure 2a shows data demonstrating the extremely high degree of 
polarization-entanglement achievable with our source.  The state 
was set to $HH - VV$; the polarization analyzer in path 1 was 
set to $-45^{\circ}$, and the other was 
varied by rotating the HWP in path 2.  As expected, the
coincidence rate displayed sinusoidal fringes with 
nearly perfect visibility ($V = 99.6 \pm 0.3\%$ with 
``accidental'' coincidences subtracted; $98.8 \pm 0.2\%$ with
them included), while the singles rate was much flatter ($V < 3.4\%$) 
\cite{fluorescence}.
\begin{figure}
\begin{center}
\epsfxsize=\columnwidth
\epsfbox{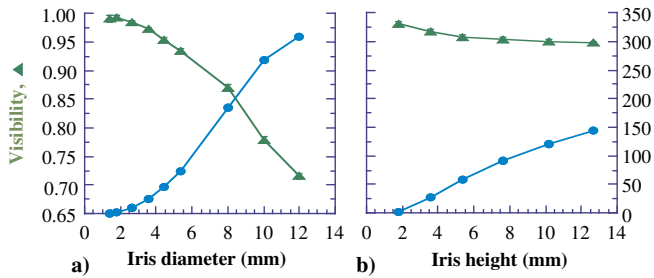}
\end{center}
\caption{
The fringe visibility (triangles, left axis) and normalized 
coincidence count rate (circles, right axis) 
versus a) the diameter of circular collection irises; and b) the vertical
opening of apertures with fixed horizontal width 3.5 mm.}
\label{fig_3}
\end{figure}
\noindent We believe this to be the highest purity 
entangled state ever reported.  The collection irises for this data 
were both only 1.76 mm in diameter -- the resulting
collection efficiency (the probability of collecting one photon 
{\it conditioned} on collecting the other) is then $\sim 10 \%$.  

To experimentally verify that we could set $\phi$ by changing 
the ellipticity of the pump light, the quarter waveplate 
(zero-order, at 351nm) before the crystals was tilted about its optic axis
(oriented vertically), thereby varying the relative phase between 
horizontal and vertical polarization components \cite{waveplate}. 
Figure 2b shows the coincidence rate with both 
analyzers at $45^{\circ}$.
For $\phi= 0,\pi$, the states 
$HH \pm VV$ are produced.  Just as with the previous 
type-II source \cite{BBOtype2}, the other two Bell states $HV\pm VH$ 
may be prepared simply by inserting a half waveplate in one of the 
arms to exchange $H$ and $V$ polarization.

To characterize the source robustness and brightness,
we measured the visibility 
as a function of the size of the collection 
apertures, located 1 m from the BBO crystals.  Opening these apertures
increases the aforementioned collection efficiency. 
In the first set of data (Fig. 3a), circular irises 
were used; the visibility decreased somewhat as the 
iris size increased, while the coincidence rate (normalized
by the input pump power) increased.  In the second set of 
measurements (Fig. 3b), a vertical slit of width 3.5mm was added
after each iris, and the vertical dimension of the aperture was
varied using the iris size; this effectively collects
a larger portion of the same cone.  The visibility then stayed 
essentially constant at $\sim 95\%$,
but the coincidence rate still increased.  At the maximum opening
(limited by our collection lens), we observed over 140
coincidences per second per milliwatt of pump power.  For 150-mW
pump power, this implies a coincidence rate of 
21,000s$^{-1}$ \cite{Buttler}, a
$\times10$ increase over the previous type-II source (which 
used a BBO crystal 2.5 times longer\cite{BBOtype2}). 
Note that this iris size  
still only accesses $\sim8\%$ of the down-conversion cone. Given
the symmetry of the arrangement, we expect strong entanglement
over the entire cone, implying a total polarization-entangled 
pair {\it production} rate (over the 5-nm bandwidth) of about
\mbox{10,000 s$^{-1}$mW$^{-1}$},
where we have divided out the filter transmissions
and detector efficiencies.
\begin{figure}
\begin{center}
\epsfxsize=\columnwidth
\epsfbox{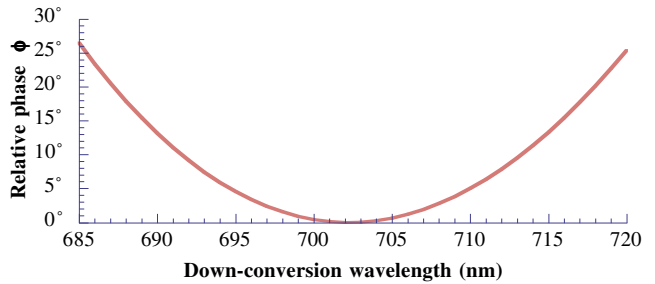}
\end{center}
\caption{A calculation of the relative phase $\phi$, as a function of the
wavelength of one of the down-conversion photons. An
overall phase offset has been suppressed for clarity.}
\label{fig_4}
\end{figure}

As a final demonstration of the source, a measurement of Bell's
inequality was performed with the 5-nm interference filters 
replaced by 10nm-wide filters (centered at 702nm), the UV
pump power increased to 60mW, and the irises set at 3.5 x 12.7 mm.  
The coincidence rates were recorded for 16 combinations of
analyzer settings [$\theta_1 =0, 90^{\circ}$, -$45^{\circ},
45^{\circ}$; $\theta_2 = -22.5^{\circ}, 67.5^{\circ},22.5^{\circ},
112.5^{\circ}$].  Following \cite{Aspect}, these may be combined to 
yield a value for the parameter $S =  2.7007 \pm 0.0029$, where 
according to any local realistic theory $|S| \le 2$ 
(and the maximum according to
quantum mechanics is $2 \sqrt{2}$).  Due to the very high
coincidence count rates obtained for this measurement, over
10,000s$^{-1}$, the necessary statistics for this 242-$\sigma$ 
violation were obtained in only 160 s!

We have thus far only considered photons belonging to a 
single cone of colors, though the arguments 
should apply to every such cone, even 
for down-converted photons with
non-degenerate frequencies. However, due to 
dispersion in the nonlinear crystals, the relative phase 
$\phi$ will in general depend on the particular wavelength
pairs being considered \cite{azimuth}. 
Fig. 4 shows the results of a 
numerical calculation of $\phi$ (modulo $360^{\circ}$), 
as a function of the wavelength of one of the 
down-conversion photons, for our particular
crystals.  
For all detected down-conversion photons to be described by 
essentially the {\it same} polarization-entangled
state, the bandwidth of acceptance needs to be
restricted, the crystal thicknesses reduced,
or a special  
birefringent compensation element included. 
We see that an acceptable range of phase 
variation ($\phi \le 26^{\circ}$, the
value for which fringe visibility $V = \cos{\phi} \ge 0.9$) 
is maintained for a bandwidth  of 30 nm,
assuming no other visibility-degrading effects come into play.  
Scaling our earlier 5-nm bandwidth result, we thus expect a
total output over the entire cones making up this
bandwidth of \mbox{$\sim 60,000$s$^{-1}$mW$^{-1}$}.
This is $\sim$300 times brighter than the polarization-entangled
photon-pair production rates obtainable with the previous
down-conversion scheme \cite{BBOtype2}, 
(and 750 times brighter if 
scaled by the crystal thickness). 

Another remarkable feature of this source is that it may be used to 
produce ``non-maximally entangled" states, i.e., states of the form 
$HH + \epsilon VV$, $|\epsilon|\ne 1$, simply by rotating the 
pump polarization -- for a pump polarized at angle 
$\theta$ to the vertical, $\epsilon = \rm{tan}\theta$.  
Such states have been shown to be useful in reducing the required 
detector efficiencies in loophole-free tests of Bell's inequalities 
\cite{Eberhard}. They are also central to certain {\it gedanken} experiments 
demonstrating the nonlocality of quantum mechanics {\it without} 
the need for inequalities \cite{Hardy}, and enlarge 
the accessible Hilbert space of quantum states.  To our knowledge, 
this source is the first one to enable preparation of such states, at 
{\it any} rate of production \cite{HardyExps}.  

Moreover, we can also 
create arbitrary 
\mbox{(partially-)} mixed 
states of the type $\rm{cos}^2\theta |H_1,H_2\protect\rangle
\langle H_2,H_1| + 
\rm{sin}^2\theta |V_1,V_2\protect\rangle
\langle V_2,V_1|$. 
We need only impose on the pump beam a 
polarization-dependent time delay which is greater than the
pump coherence time (for mixed states) or comparable to it 
(for partially-mixed states)\cite{duality}. 

Finally, as indicated earlier, the down-conversion photon pairs are 
automatically entangled in energy and momentum 
as well.  Hence, for our two-crystal scheme, the photons are 
actually simultaneously entangled in {\it all} degrees of freedom. We call 
such a state ``hyper-entangled" \cite{hyper}, and it has been 
shown that such states may benefit certain experiments 
in quantum information \cite{Strekalov,embedded}.   

In summary, using spontaneous down-conversion in a very simple 
two-crystal geometry, we have demonstrated a tunable source of 
polarization-entangled photon pairs. Because the entanglement 
exists over the entire cones of emitted light, 
this source is much
brighter than previous ones, allowing a tremendous 
Bell inequality violation in only minutes.  
Such brightness is completely necessary for some applications 
(like quantum cryptography to a satellite), and
very advantageous for others (like teleportation, 
which requires two pairs of entangled photons,
and hence scales as the {\it square} of the source intensity). 
Due to its simplicity 
and robustness, this source should benefit many  
ongoing pursuits using correlated photons pairs, and 
may even permit the inclusion of tests of nonlocality  
in standard undergraduate physics labs. 

Current addresses: 
$^{*}$ Ginzton Laboratory, Stanford Univ., Stanford, CA 94305;
$^{\dagger}$ Physics Dept., M.I.T., Cambridge, MA 02139


\begin{references}
\bibitem{overlap} Based on a simple geometrical argument,
the spatial overlap will be high if 
$\theta_{dc}L/D \ll 1$, with cone opening angle $\theta_{dc}$, 
crystal thickness $L$, and pump beam diamter $D$.  
The constraint on transverse walkoff in the crystals is 
similar if we replace $\theta_{dc}$ by the walkoff angle $\rho$ 
(typically $\approx 4^{\circ}$).
\bibitem{fluorescence}
Some of the singles rate net polarization may be due to an 
unexpectedly high background fluorescence from the BBO crystals,  
nearly one third of the true singles rate.
\end{references}
\end{document}